\documentclass[11pt,letter]{article}
\usepackage{deauthor}
\usepackage{times}
\usepackage{graphicx}
\usepackage{amsfonts}   
\usepackage{amsmath}
\usepackage{amssymb}
\usepackage{booktabs}
\usepackage{multirow}
\usepackage[font=small,labelfont=bf]{caption}
\usepackage[linesnumbered,ruled]{algorithm2e}

\newcommand{\xhdr}[1]{{\noindent\bfseries #1}.}

\newcommand{\mb}{\mathbf}
\newcommand{\G}{\mathcal{G}}
\newcommand{\V}{\mathcal{V}}
\newcommand{\E}{\mathcal{E}}
\newcommand{\R}{\mathbb{R}}

\newcommand{\enc}{\textsc{enc}}
\newcommand{\dec}{\textsc{dec}}

\begin{document}

\title{Representation Learning on Graphs: Methods and Applications}
\def\sharedaffiliation{%
\end{tabular}
\begin{tabular}{c}}
\author{
William L. Hamilton\\
\texttt{wleif@stanford.edu}
\and
Rex Ying\\
\texttt{rexying@stanford.edu}
\and
Jure Leskovec \\
\texttt{jure@cs.stanford.edu}
\vspace{10pt}
\sharedaffiliation
Department of Computer Science\\
Stanford University\\
Stanford, CA, 94305
}

\maketitle

\begin{abstract}
Machine learning on graphs is an important and ubiquitous task with  applications ranging from drug design to friendship recommendation in social networks.
The primary challenge in this domain is finding a way to represent, or encode, graph structure so that it can be easily exploited by machine learning models.
Traditionally, machine learning approaches relied on user-defined heuristics to extract features encoding structural information about a graph (e.g., degree statistics or kernel functions). 
However, recent years have seen a surge in approaches that automatically learn to encode graph structure into low-dimensional embeddings, using techniques based on deep learning and nonlinear dimensionality reduction. 
Here we provide a conceptual review of key advancements in this area of representation learning on graphs, including matrix factorization-based methods, random-walk based algorithms, and graph neural networks. 
We review methods to embed individual nodes as well as approaches to embed entire (sub)graphs.
In doing so, we develop a unified framework to describe these recent approaches, and we highlight a number of important applications and directions for future work. 
\end{abstract}

\section{Introduction}

Graphs are a ubiquitous data structure, employed extensively within  computer science and related fields. 
Social networks, molecular graph structures, biological protein-protein networks, recommender systems---all of these domains and many more can be readily modeled as graphs, which capture interactions (\textit{i.e.}, edges) between individual units (\textit{i.e.}, nodes). 
As a consequence of their ubiquity, graphs are the backbone of countless systems, allowing relational knowledge about interacting entities to be efficiently stored and accessed \cite{angles2008survey}.

However, graphs are not only useful as structured knowledge repositories: they also play a key role in modern machine learning. 
Many machine learning applications seek to make predictions or discover new patterns using graph-structured data as feature information.
For example, one might wish to classify the role of a protein in a biological interaction graph, predict the role of a person in a collaboration network, recommend new friends to a user in a social network, or predict new therapeutic applications of existing drug molecules, whose structure can be represented as a graph.

\begin{figure}
\centering
\includegraphics[width=0.9\textwidth, bb=0 0 500 200]{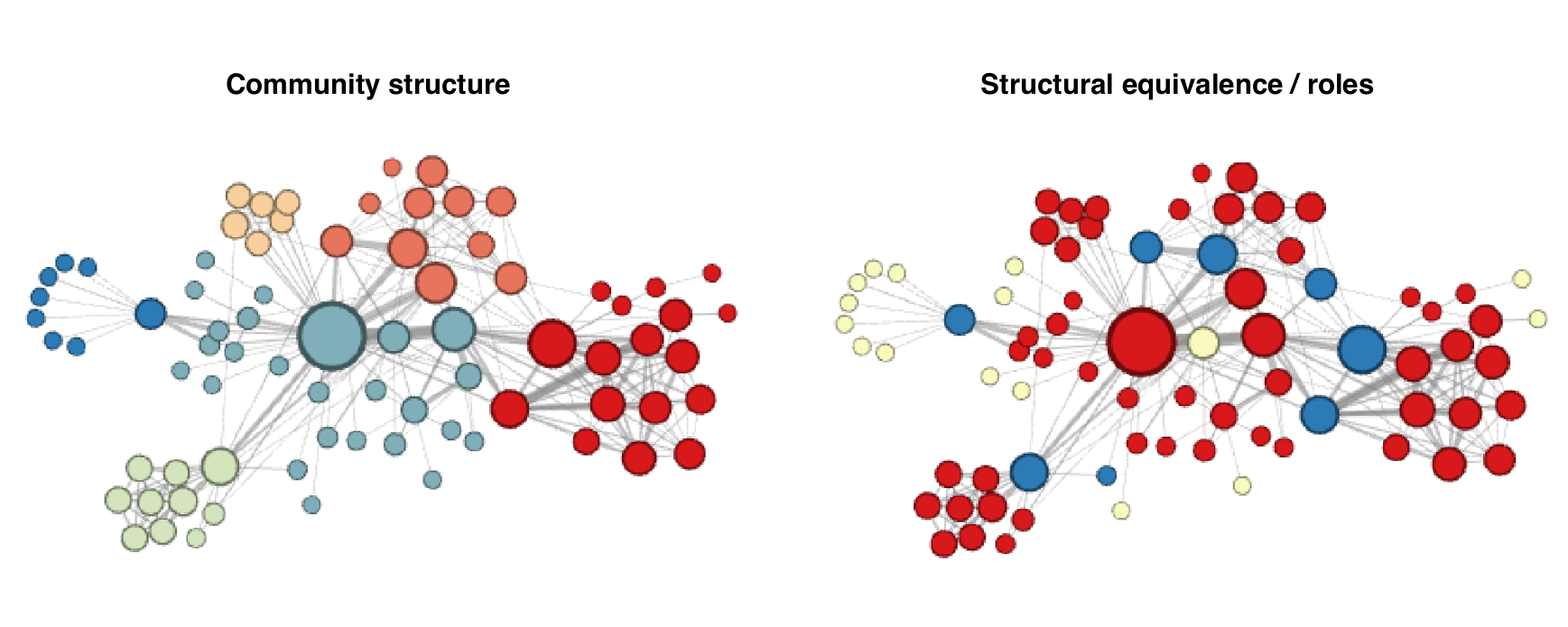}
\caption{
Two different views of a character-character interaction graph derived from the Les Mis\'erables novel, where two nodes are connected if the corresponding characters interact. 
The coloring in the left figure emphasizes differences in the nodes' global positions in the graph: nodes have the same color if they belong to the same community, at a global level. 
In contrast, the coloring in the right figure denotes structural equivalence between nodes, or the fact that two nodes play similar roles in their local neighborhoods (\textit{e.g.}, ``bridging nodes'' are colored blue).
The colorings for both figures were generated using different settings of the node2vec node embedding method \cite{grover2016node2vec}, described in Section \ref{sec:nodes}.  Reprinted from \cite{grover2016node2vec} with permission.\protect\footnotemark}
\label{fig:communityvsroles}
\end{figure}

The central problem in machine learning on graphs is finding a way to incorporate information about graph-structure into a machine learning model. 
For example, in the case of link prediction in a social network, one might want to encode pairwise properties between nodes, such as relationship strength or the number of common friends.
Or in the case of node classification, one might want to include information about the global position of a node in the graph or the structure of the node's local graph neighborhood (Figure \ref{fig:communityvsroles}).
The challenge---from a machine learning perspective---is that there is no straightforward way to encode this high-dimensional, non-Euclidean information about graph structure into a feature vector. 

To extract structural information from graphs, traditional machine approaches often rely on summary graph statistics (\textit{e.g.}, degrees or clustering coefficients) \cite{bhagat2011node},  kernel functions \cite{vishwanathan2010graph}, or carefully engineered features to measure local neighborhood structures \cite{liben2007link}.
However, these approaches are limited because these hand-engineered features are inflexible---\textit{i.e.}, they cannot adapt during the learning process---and designing these features can be a time-consuming and expensive process. 

More recently, there has been a surge of approaches that seek to {\em learn} representations that encode structural information about the graph.
The idea behind these {\em representation learning} approaches is to learn a mapping that embeds nodes, or entire (sub)graphs, as points in a low-dimensional vector space $\R^d$.
The goal is to optimize this mapping so that geometric relationships in the embedding space reflect the structure of the original graph.
After optimizing the embedding space, the learned embeddings can be used as feature inputs for downstream machine learning tasks.  
The key distinction between representation learning approaches and previous work is how they treat the problem of representing graph structure.
Previous work treated this problem as a pre-processing step, using hand-engineered statistics to extract structural information.
In contrast, representation learning approaches treat this problem as machine learning task itself, using a data-driven approach to learn embeddings that encode graph structure. 

\footnotetext{For this and all subsequent reprinted figures, the original authors retain their copyrights, and permission was obtained from the corresponding author.}
Here we provide an overview of recent advancements in representation learning on graphs, reviewing techniques for representing both nodes and entire subgraphs. 
Our survey attempts to merge together multiple, disparate lines of research that have drawn significant attention across different subfields and venues in recent years---\textit{e.g.}, node embedding methods, which are a popular object of study in the data mining community,  
and graph convolutional networks, which have drawn considerable attention in major machine learning venues. 
In doing so, we develop a unified conceptual framework for describing the various approaches and emphasize major conceptual distinctions.

We focus our review on recent approaches that have garnered significant attention in the machine learning and data mining communities, especially methods that are scalable to massive graphs (\textit{e.g.}, millions of nodes) and inspired by advancements in deep learning. 
Of course, there are other lines of closely related and relevant work, which we do not review in detail here---including latent space models of social networks \cite{hoff2002latent}, embedding methods for statistical relational learning \cite{nickel2016review}, manifold learning algorithms \cite{lee2007nonlinear}, and geometric deep learning \cite{bronstein2017geometric}---all of which involve representation learning with graph-structured data. 
We refer the reader to \cite{hoff2002latent}, \cite{nickel2016review}, \cite{lee2007nonlinear}, and \cite{bronstein2017geometric} for comprehensive overviews of these areas.

\subsection{Notation and essential assumptions}
We will assume that the primary input to our representation learning algorithm is an undirected graph $\G = (\V, \E)$ with associated binary adjacency matrix $\mb{A}$.\footnote{Most of the methods we review are easily generalized to work with weighted or directed graphs, and we will explicitly describe how to generalize certain methods to the multi-modal setting (\textit{i.e.}, differing node and edge types).}
We also assume that the methods can make use of a real-valued matrix of node attributes $\mb{X} \in \R^{m \times |\V|}$ (\textit{e.g.}, representing text or metadata associated with nodes).
The goal is to use the information contained in $\mb{A}$ and $\mb{X}$ to map each node, or a subgraph, to a vector $\mb{z} \in \R^d$, where $d << |\V|$. 

Most of the methods we review will optimize this mapping in an {\em unsupervised} manner, making use of only information in $\mb{A}$ and $\mb{X}$, without knowledge of a particular downstream machine learning task.
However, we will also discuss some approaches for {\em supervised} representation learning, where the models make use of classification or regression labels in order to optimize the embeddings. 
These classification labels may be associated with individual nodes or entire subgraphs and are the prediction targets for downstream machine learning tasks (\textit{e.g.}, they might label protein roles, or the therapeutic properties of a molecule, based on its graph representation).  

\begin{figure}
\centering
\includegraphics[width=0.9\textwidth, bb=0 0 1024 345]{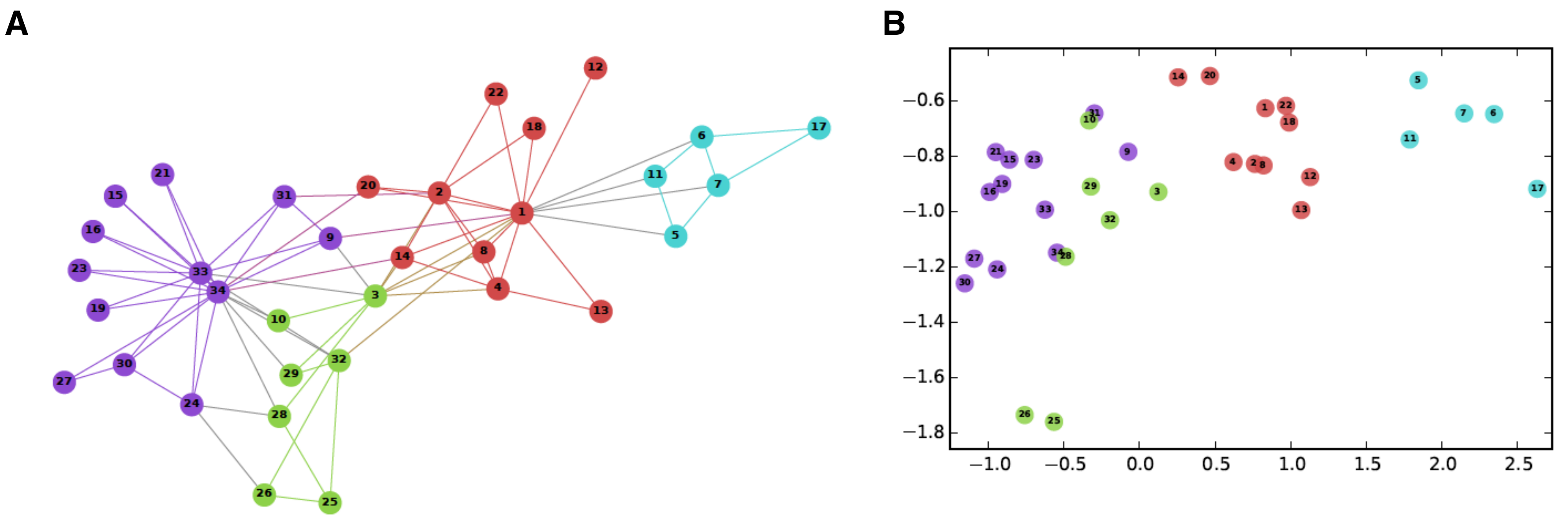}
\caption{
\textbf{A}, Graph structure of the Zachary Karate Club social network, where nodes are connected if the corresponding individuals are friends. 
The nodes are colored according to the different communities that exist in the network.
\textbf{B}, Two-dimensional visualization of node embeddings generated from this graph using the DeepWalk method (Section \ref{sec:randwalk}) \cite{perozzi2014deepwalk}.
The distances between nodes in the embedding space reflect similarity in the original graph, and the node embeddings are spatially clustered according to the different color-coded communities. 
Reprinted with permission from \cite{perozzi2014deepwalk,perozzithesis}.}
\label{fig:karate}
\end{figure}

\section{Embedding nodes}\label{sec:nodes}

We begin with a discussion of methods for {\em node embedding}, where the goal is to encode nodes as low-dimensional vectors that summarize their graph position and the structure of their local graph neighborhood.
These low-dimensional embeddings can be viewed as encoding, or projecting, nodes into a latent space, where geometric relations in this latent space correspond to interactions (\textit{e.g.}, edges) in the original graph \cite{hoff2002latent}.
Figure \ref{fig:karate} visualizes an example embedding of the famous Zachary Karate Club social network \cite{perozzi2014deepwalk}, where two dimensional node embeddings capture the community structure implicit in the social network.

\subsection{Overview of approaches: An encoder-decoder perspective}\label{sec:encdec}

\begin{figure}
\centering
\includegraphics[width=0.8\textwidth, bb=0 0 1024 350]{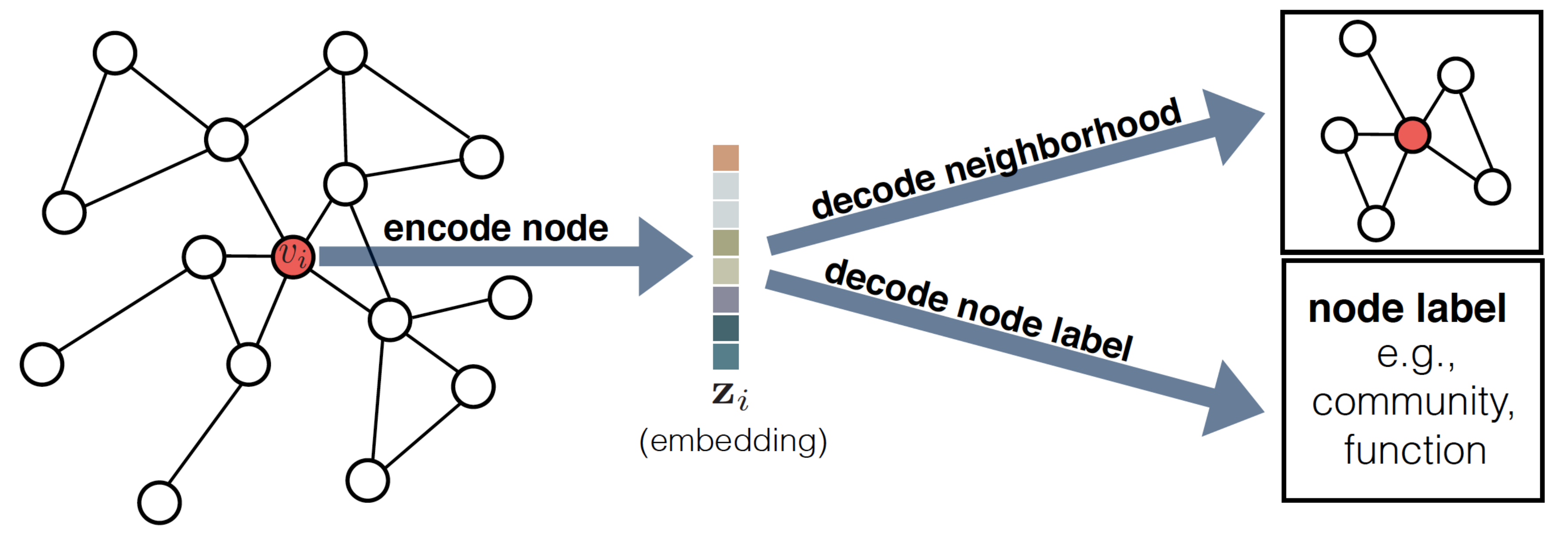}
\caption{Overview of the encoder-decoder approach. First the encoder maps the node, $v_i$, to a low-dimensional vector embedding, $\mb{z}_i$, based on the node's position in the graph, its local neighborhood structure, and/or its attributes. Next, the decoder extracts user-specified information from the low-dimensional embedding; this might be information about $v_i$'s local graph neighborhood (\textit{e.g.}, the identity of its neighbors) or a classification label associated with $v_i$ (\textit{e.g.}, a community label).  By jointly optimizing the encoder and decoder, the system learns to compress information about graph structure into the low-dimensional embedding space.}
\label{fig:encdec}
\end{figure}

Recent years have seen a surge of research on node embeddings, leading to a complicated diversity of notations, motivations, and conceptual models. 
Thus, before discussing the various techniques, we first develop a unified {\em encoder-decoder} framework, which explicitly structures this methodological diversity and puts the various methods on equal notational and conceptual footing.

In this framework, we organize the various methods around two key mapping functions: an {\em encoder}, which maps each node to a low-dimensional vector, or embedding, and a {\em decoder}, which decodes structural information about the graph from the learned embeddings (Figure \ref{fig:encdec}). 
The intuition behind the encoder-decoder idea is the following: if we can learn to decode high-dimensional graph information---such as the global positions of nodes in the graph or the structure of local graph neighborhoods---from encoded low-dimensional embeddings, then, in principle, these embeddings should contain all information necessary for downstream machine learning tasks.

Formally, the {\em encoder} is a function,
\begin{equation}\label{eq:enc}
\enc : \mathcal{V} \rightarrow \mathbb{R}^d,
\end{equation}
that maps nodes to vector embeddings $\mb{z}_i \in \mathbb{R}^{d}$ (where $\mb{z}_i$ corresponds to the embedding for node $v_i \in \V$).
The {\em decoder} is a function that accepts a set of node embeddings and decodes user-specified graph statistics from these embeddings.
For example, the decoder might predict the existence of edges between nodes, given their embeddings \cite{ahmed2013distributed,kipf2016variational}, or it might predict the community that a node belongs to in the graph \cite{hamilton2017inductive,kipf2016semi} (Figure \ref{fig:encdec}).  
In principle, many decoders are possible; however, the vast majority of works use a basic {\em pairwise decoder},
\begin{equation}\label{eq:dec}
\dec : \mathbb{R}^d \times \mathbb{R}^d \rightarrow \mathbb{R}^+,
\end{equation}
that maps pairs of node embeddings to a real-valued node similarity measure, which quantifies the similarity of the two nodes in the original graph.

When we apply the pairwise decoder to a pair of embeddings ($\mb{z}_i$,$\mb{z}_j$) we get a {\em reconstruction} of the similarity between $v_i$ and $v_j$ in the original graph, and the goal is optimize the encoder and decoder mappings to minimize the error, or loss, in this reconstruction so that:
\begin{align}\label{eq:reconstruction}
\dec(\enc(v_i), \enc(v_j)) &= \dec(\mb{z}_i, \mb{z}_j) \approx s_{\G}(v_i, v_j),
\end{align}
where $s_{\G}$ is a user-defined, graph-based similarity measure between nodes, defined over the graph $\G$. 
In other words, we want to optimize our encoder-decoder model so that we can decode pairwise node similarities in the original graph $s_{\G}(v_i, v_j)$ from the low-dimensional node embeddings $\mb{z}_i$ and $\mb{z}_j$.
For example, one might set $s_{\G}(v_i, v_j) \triangleq \mb{A}_{i,j}$ and define nodes to have a similarity of 1 if they are adjacent and 0 otherwise \cite{ahmed2013distributed}, or one might define $s_G$ according to the probability of $v_i$ and $v_j$ co-occurring on a fixed-length random walk over the graph $\G$ \cite{grover2016node2vec,perozzi2014deepwalk}.
In practice, most approaches realize the reconstruction objective (Equation \ref{eq:reconstruction}) by minimizing an empirical loss $\mathcal{L}$ over a set of training node pairs $\mathcal{D}$:
\begin{equation}\label{eq:loss}
\mathcal{L} = \sum_{(v_i, v_j) \in \mathcal{D}}\ell\left(\dec(\mb{z}_i, \mb{z}_j), s_\G(v_i, v_j)\right),
\end{equation}
where $\ell \::\: \R \times \R \rightarrow \R$ is a user-specified loss function, which measures the discrepancy between the decoded (\textit{i.e.}, estimated) similarity values $\dec(\mb{z}_i, \mb{z}_j)$ and the true values $s_\G(v_i, v_j)$.

Once we have optimized the encoder-decoder system, we can use the trained encoder to generate embeddings for nodes, which can then be used as a feature inputs for downstream machine learning tasks. 
For example, one could feed the learned embeddings to a logistic regression classifier to predict the community that a node belongs to \cite{perozzi2014deepwalk}, or one could use distances between the embeddings to recommend friendship links in a social network \cite{backstrom2011supervised,grover2016node2vec} (Section \ref{sec:nodeapplications} discusses further applications).

Adopting this encoder-decoder view, we organize our discussion of the various node embedding methods along the following four methodological components:
\begin{enumerate}
\item
	A {\bf pairwise similarity function} $s_\G : \V \times \V \rightarrow \R^{+}$, defined over the graph $\G$. 
	This function measures the similarity between nodes in $\G$.
\item
	An {\bf encoder function}, $\enc$, that generates the node embeddings. This function contains a number of trainable parameters that are optimized during the training phase. 
\item
	A {\bf decoder function}, $\dec$, which reconstructs pairwise similarity values from the generated embeddings.  This function usually contains no trainable parameters.
\item
 A {\bf loss function}, $\ell$, which determines how the quality of the pairwise reconstructions is evaluated in order to train the model, \textit{i.e.}, how $\dec(\mb{z}_i, \mb{z}_j)$ is compared to the true $s_\G(v_i, v_j)$ values.
\end{enumerate}
As we will show, the primary methodological distinctions between the various node embedding approaches are in how they define these four components.

\subsubsection{Notes on optimization and implementation details}

All of the methods we review involve optimizing the parameters of the encoder algorithm, $\Theta_{\enc}$, by minimizing a loss analogous to Equation \eqref{eq:loss}.\footnote{Occasionally, different methods will add additional auxiliary objectives or regularizers beyond the standard encoder-decoder objective, but we will often omit these details for brevity. A few methods also optimize parameters in the decoder, $\Theta_\dec$.}
In most cases, stochastic gradient descent is used for optimization, though some algorithms do permit closed-form solutions via matrix decomposition (\textit{e.g.}, \cite{cao2015grarep}).  
However, note that we will not focus on optimization algorithms here and instead will emphasize high-level differences that exist across different embedding methods, independent of the specifics of the optimization approach. 

\subsection{Shallow embedding approaches}\label{sec:directencoding}

The majority of node embedding algorithms rely on what we call {\em shallow embedding}.
For these shallow embedding approaches, the encoder function---which maps nodes to vector embeddings---is simply an ``embedding lookup'':
\begin{equation}\label{eq:embedlookup}
\enc(v_i) = \mb{Z}\mb{v}_i,
\end{equation}
where $\mb{Z} \in \mathbb{R}^{d \times |\V|}$ is a matrix containing the embedding vectors for all nodes and $\mb{v}_i \in \mathbb{I}_{\V}$ is a one-hot indicator vector indicating the column of $\mb{Z}$ corresponding to node $v_i$. 
The set of trainable parameters for shallow embedding methods is simply $\Theta_{\enc} = \{\mb{Z}\}$, \textit{i.e.}, the embedding matrix $\mb{Z}$ is optimized directly. 

These approaches are largely inspired by classic matrix factorization techniques for dimensionality reduction \cite{belkin2002laplacian} and multi-dimensional scaling \cite{kruskal1964multidimensional}.
Indeed, many of these approaches were originally motivated as factorization algorithms, and we reinterpret them within the encoder-decoder framework here. 
Table \ref{tab:directencoding} summarizes some well-known shallow embedding methods within the encoder-decoder framework. 
Table \ref{tab:directencoding} highlights how these methods can be succinctly described according to (i) their decoder function, (ii) their graph-based similarity measure, and (iii) their loss function.  
The following two sections describe these methods in more detail, distinguishing between matrix factorization-based approaches (Section \ref{sec:fact}) and more recent approaches based on random walks (Section \ref{sec:randwalk}). 

\begin{table}
\caption{A summary of some well-known shallow embedding embedding algorithms. Note that the decoders and similarity functions for the random-walk based methods are asymmetric, with the similarity function $p_{\G}(v_j | v_i)$ corresponding to the probability of visiting $v_j$ on a fixed-length random walk starting from $v_i$.}
\label{tab:directencoding}
\centering
\begin{tabular}{cccccc}
\toprule
Type & Method & Decoder &  Similarity measure & Loss function ($\ell$)\\
\midrule
& {\small Laplacian Eigenmaps} \cite{belkin2002laplacian} & $\|\mb{z}_i - \mb{z}_j\|_2^2$ & general & $\dec(\mb{z}_i, \mb{z}_j)\cdot s_\G(v_i,v_j)$\\
Matrix & {\small Graph Factorization} \cite{ahmed2013distributed} & $\mb{z}_i^\top\mb{z}_j$ & $\mb{A}_{i,j}$ & $\|\dec(\mb{z}_i, \mb{z}_j) - s_\G(v_i,v_j)\|_2^2$\\
factorization & {\small GraRep} \cite{cao2015grarep} & $\mb{z}_i^\top\mb{z}_j$ & $\mb{A}_{i,j}, \mb{A}^2_{i,j}, ..., \mb{A}^k_{i,j}$ & $\|\dec(\mb{z}_i, \mb{z}_j) - s_\G(v_i,v_j)\|_2^2$\\
& {\small HOPE} \cite{ou2016asymmetric} & $\mb{z}_i^\top\mb{z}_j$ & general & $\|\dec(\mb{z}_i, \mb{z}_j) - s_\G(v_i,v_j)\|_2^2$\\
\midrule
\multirow{3}{*}{Random walk}
 & {\small DeepWalk} \cite{perozzi2014deepwalk} & $\frac{e^{\mb{z}_i^\top\mb{z}_j}}{\sum_{k\in \V}e^{\mb{z}_i^\top\mb{z}_k}}$ & $p_\G(v_j | v_i)$ & $-s_\G(v_i, v_j)\log(\dec(\mb{z}_i, \mb{z}_j))$ \vspace{3pt} \\
  & {\small node2vec} \cite{grover2016node2vec} & $\frac{e^{\mb{z}_i^\top\mb{z}_j}}{\sum_{k\in \V}e^{\mb{z}_i^\top\mb{z}_k}}$ & $p_\G(v_j | v_i)$  (biased)& $-s_\G(v_i, v_j)\log(\dec(\mb{z}_i, \mb{z}_j))$
\end{tabular}
\end{table}

\subsubsection{Factorization-based approaches}\label{sec:fact}

Early methods for learning representations for nodes largely focused on matrix-factorization approaches, which are directly inspired by classic techniques for dimensionality reduction \cite{belkin2002laplacian,kruskal1964multidimensional}.

\xhdr{Laplacian eigenmaps} One of the earliest, and most well-known instances, is the Laplacian eigenmaps (LE) technique \cite{belkin2002laplacian}, which we can view within the encoder-decoder framework as a shallow embedding approach in which the decoder is defined as
$$\dec(\mb{z}_i, \mb{z}_j) = \|\mb{z}_i-\mb{z}_j\|_2^2$$ and
where the loss function weights pairs of nodes according to their similarity in the graph:
\begin{equation}\label{eq:le}
\mathcal{L} = \sum_{(v_i, v_j) \in \mathcal{D}} \dec(\mb{z}_i, \mb{z}_j)\cdot s_\G(v_i, v_j).
\end{equation}

\xhdr{Inner-product methods} Following on the Laplacian eigenmaps technique, there are a large number of recent embedding methodologies based on a pairwise, inner-product decoder: 
\begin{equation}\label{eq:innerprod}
\dec(\mb{z}_i, \mb{z}_j) = \mb{z}_i^\top\mb{z}_j,
\end{equation}
where the strength of the relationship between two nodes is proportional to the dot product of their embeddings.
The Graph Factorization (GF) algorithm\footnote{Of course, Ahmed et al. \cite{ahmed2013distributed} were not the first researchers to propose factorizing an adjacency matrix, but they were the first to present a scalable $O(|\E|)$ algorithm for the purpose of generating node embeddings.} \cite{ahmed2013distributed}, GraRep \cite{cao2015grarep}, and HOPE \cite{ou2016asymmetric} all fall firmly within this class.
In particular, all three of these methods use an inner-product decoder, a mean-squared-error (MSE) loss, 
\begin{equation}\label{eq:factloss}
\mathcal{L} = \sum_{(v_i, v_j) \in \mathcal{D}}\|\dec(\mb{z}_i, \mb{z}_j) - s_\G(v_i,v_j)\|_2^2,
\end{equation}
and they differ primarily in the node similarity measure used, \textit{i.e.} how they define $s_\G(v_i, v_j)$.
The Graph Factorization algorithm defines node similarity directly based on the adjacency matrix (\textit{i.e.}, $s_\G(v_i, v_j) \triangleq \mb{A}_{i,j}$) \cite{ahmed2013distributed}; GraRep considers various powers of the adjacency matrix (\textit{e.g.}, $s_\G(v_i, v_j) \triangleq \mb{A}_{i,j}^2$) in order to capture higher-order node similarity \cite{cao2015grarep}; and the HOPE algorithm supports general similarity measures (\textit{e.g.}, based on Jaccard neighborhood overlaps) \cite{ou2016asymmetric}.
These various different similarity functions trade-off between modeling ``first-order similarity'', where $s_\G$ directly measures connections between nodes (\textit{i.e.}, $s_\G(v_i, v_j) \triangleq \mb{A}_{i,j}$ \cite{ahmed2013distributed}) and modeling ``higher-order similarity'', where $s_\G$ corresponds to more general notions of neighborhood overlap (\textit{e.g.}, $s_\G (v_i, v_j)= \mb{A}^2_{i,j}$ \cite{cao2015grarep}).

We refer to these methods in this section as matrix-factorization approaches because, averaging over all nodes, they optimize loss functions (roughly) of the form:
\begin{equation}
\mathcal{L} \approx \|\mb{Z}^\top\mb{Z} - \mb{S}\| _2^2,
\end{equation}
where $\mb{S}$ is a matrix containing pairwise similarity measures (\textit{i.e.}, $\mb{S}_{i,j} \triangleq s_\G(v_i, v_j)$) and $\mb{Z}$ is the matrix of node embeddings.
Intuitively, the goal of these methods is simply to learn embeddings for each node such that the inner product between the learned embedding vectors approximates some deterministic measure of node similarity. 

\subsubsection{Random walk approaches}\label{sec:randwalk}

\begin{figure}
\centering
\includegraphics[width=0.9\textwidth, bb=0 0 690 210]{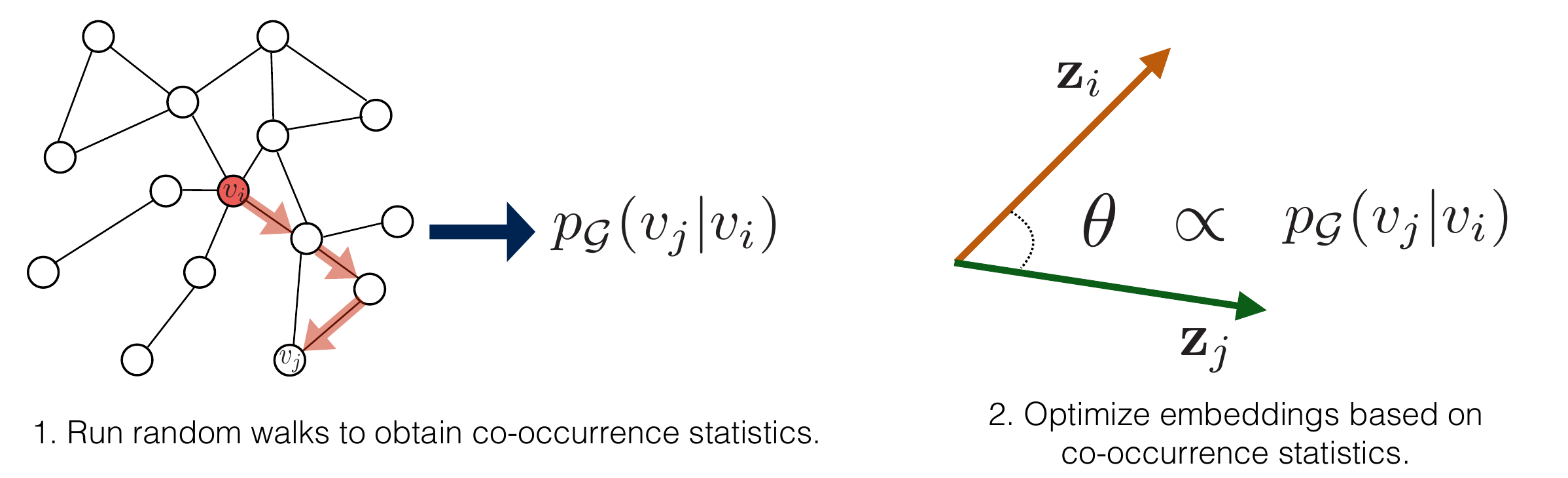}
\caption{The random-walk based methods sample a large number of fixed-length random walks starting from each node, $v_i$.
The embedding vectors are then optimized so that the dot-product, or angle, between two embeddings, $\mb{z}_i$ and $\mb{z}_j$, is (roughly) proportional to the probability of visiting $v_j$ on a fixed-length random walk starting from $v_i$.
 }
\label{fig:randwalk}
\end{figure}

Many recent successful methods that also belong to the class of shallow embedding approaches learn the node embeddings based on random walk statistics. 
Their key innovation is optimizing the node embeddings so that nodes have similar embeddings if they tend to co-occur on short random walks over the graph (Figure \ref{fig:randwalk}). 
Thus, instead of using a deterministic measure of node similarity, like the methods of Section \ref{sec:fact}, these random walk methods employ a flexible, stochastic measure of node similarity, which has led to superior performance in a number of settings \cite{goyal2017graph}.

\xhdr{DeepWalk and node2vec} 
Like the matrix factorization approaches described above, DeepWalk and node2vec rely on shallow embedding and use a decoder based on the inner product.
However, instead of trying to decode a deterministic node similarity measure, these approaches optimize embeddings to encode the statistics of random walks. 
The basic idea behind these approaches is to learn embeddings so that (roughly):
\begin{align}\label{eq:random}
\dec(\mb{z}_i, \mb{z}_j) &\triangleq \frac{e^{\mb{z}_i^\top\mb{z}_j}}{\sum_{v_k \in \V}e^{\mb{z}_i^\top\mb{z}_k}}\\\nonumber
 &\approx p_{\G, T}(v_j | v_i),
\end{align}
where $p_{\G,T}(v_j | v_i)$ is the probability of visiting $v_j$ on a length-$T$ random walk starting at $v_i$, with $T$ usually defined to be in the range $T \in \{2,...,10\}$.
Note that unlike the similarity measures in Section \ref{sec:fact}, $p_{\G,T}(v_j | v_i)$ is both stochastic and asymmetric. 

More formally, these approaches attempt to minimize the following cross-entropy loss:
\begin{equation}\label{eq:randloss}
\mathcal{L} = \sum_{(v_i, v_j) \in \mathcal{D}}-\log(\dec(\mb{z}_i, \mb{z}_j)),
\end{equation}
where the training set $\mathcal{D}$ is generated by sampling random walks starting from each node (\textit{i.e.}, where $N$ pairs for each node $v_i$ are sampled from the distribution $(v_i, v_j) \sim  p_{\G,T}(v_j | v_j)$). 
However, naively evaluating the loss in Equation \eqref{eq:randloss} is prohibitively expensive---in particular, $O(|\mathcal{D}||\V|)$---since evaluating the denominator of Equation \eqref{eq:random} has time complexity $O(|\V|)$.
Thus, DeepWalk and node2vec use different optimizations and approximations to compute the loss in Equation \eqref{eq:randloss}.
DeepWalk employs a ``hierarchical softmax'' technique to compute the normalizing factor, using a binary-tree structure to accelerate the computation \cite{perozzi2014deepwalk}.
In contrast, node2vec approximates Equation \eqref{eq:randloss} using ``negative sampling'': instead of normalizing over the full vertex set, node2vec approximates the normalizing factor using a set of random ``negative samples'' \cite{grover2016node2vec}.

\begin{figure}
\centering
\includegraphics[width=0.95\textwidth, bb=0 0 800 200]{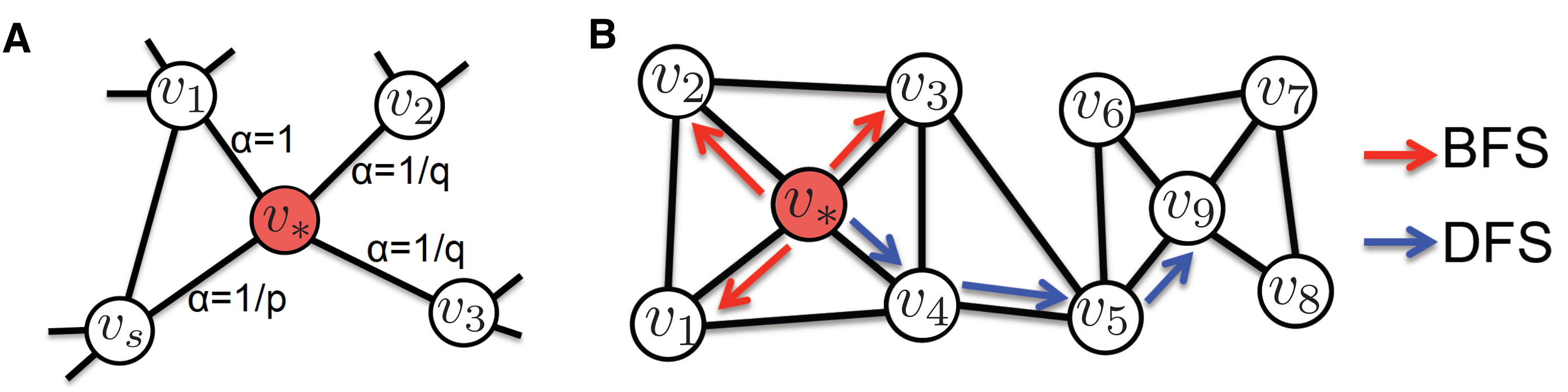}
\caption{
\textbf{A}, Illustration of how node2vec biases the random walk using the $p$ and $q$ parameters. 
Assuming that the walk just transitioned from $v_s$ to $v_*$, the edge labels, $\alpha$, are proportional to the probability of the walk taking that edge at next time-step. 
\textbf{B}, Difference between random-walks that are based on breadth-first search (BFS) and depth-first search (DFS). 
BFS-like random walks are mainly limited to exploring a node's immediate (\textit{i.e.}, one-hop) neighborhood and are generally more effective for capturing structural roles. DFS-like walks explore further away from the node and are more effective for capturing community structures. Adapted from \cite{grover2016node2vec}.}
\label{fig:n2v_walk}
\end{figure}

Beyond these algorithmic differences, the key distinction between node2vec and DeepWalk is that node2vec allows for a flexible definition of random walks, whereas DeepWalk uses simple unbiased random walks over the graph.
In particular, node2vec introduces two random walk hyperparameters, $p$ and $q$, that bias the random walk (Figure \ref{fig:n2v_walk}.A).
The hyperparameter $p$ controls the likelihood of the walk immediately revisiting a node, while $q$ controls the likelihood of the walk revisiting a node's one-hop neighborhood. 
By introducing these hyperparameters, node2vec is able to smoothly interpolate between walks that are more akin to breadth-first or depth-first search (Figure \ref{fig:n2v_walk}.B).
 Grover et al. found that tuning these parameters allowed the model to trade off between learning embeddings that emphasize community structures or embeddings that emphasize local structural roles \cite{grover2016node2vec} (see also Figure \ref{fig:communityvsroles}).

\xhdr{Large-scale information network embeddings (LINE)} Another highly successful shallow embedding approach, which is not based random walks but is contemporaneous and often compared with DeepWalk and node2vec, is the LINE method  \cite{tang2015line}.
LINE combines two encoder-decoder objectives that optimize ``first-order'' and ``second-order'' node similarity, respectively.
The first-order objective uses a decoder based on the sigmoid function,
\begin{equation}
\dec(\mb{z}_i, \mb{z}_j) = \frac{1}{1 + e^{-\mb{z}_i^\top\mb{z}_j}},
\end{equation}
and an adjacency-based similarity measure (\textit{i.e.}, $s_\G(v_i, v_j) = \mb{A}_{i,j}$).
The second-order encoder-decoder objective is similar but considers two-hop adjacency neighborhoods and uses an encoder identical to Equation \eqref{eq:random}.
Both the first-order and second-order objectives are optimized using loss functions derived from the KL-divergence metric \cite{tang2015line}.
Thus, LINE is conceptually related to node2vec and DeepWalk in that it uses a probabilistic decoder and loss, but it explicitly factorizes first- and second-order similarities, instead of combining them in fixed-length random walks. 

\xhdr{HARP: Extending random-walk embeddings via graph pre-processing}
Recently, Chen et al.\@ \cite{chen2017harp} introduced a ``meta-strategy'', called HARP, for improving various random-walk approaches via a graph pre-processing step.   
In this approach, a graph coarsening procedure is used to collapse related nodes in $\G$ together into ``supernodes'', and then DeepWalk, node2vec, or LINE is run on this coarsened graph. 
After embedding the coarsened version of $\G$, the learned embedding of each supernode is used as an initial value for the random walk embeddings of the supernode's constituent nodes (in another round of optimization on a ``finer-grained'' version of the graph).
This general process can be repeated in a hierarchical manner at varying levels of coarseness, and has been shown to consistently improve performance of DeepWalk, node2vec, and LINE \cite{chen2017harp}.

\xhdr{Additional variants of the random-walk idea} There have also been a number of further extensions of the random walk idea.
For example, Perozzi et al. \cite{perozzi2016walklets} extend the DeepWalk algorithm to learn embeddings using random walks that ``skip'' or ``hop'' over multiple nodes at each step, resulting in a similarity measure similar to GraRep \cite{cao2015grarep}, while Chamberlan et al.\@ \cite{chamberlain2017neural} modify the inner-product decoder of node2vec to use a hyperbolic, rather than Euclidean, distance measure.

\subsection{Generalized encoder-decoder architectures}\label{sec:genencdec}

So far all of the node embedding methods we have reviewed have been shallow embedding methods, where the encoder is a simply an embedding lookup (Equation \ref{eq:embedlookup}). 
However, these shallow embedding approaches train unique embedding vectors for each node independently, which leads to a number of drawbacks:
\begin{enumerate}
\item
	No parameters are shared between nodes in the encoder (\textit{i.e.}, the encoder is simply an embedding lookup based on arbitrary node ids).  
 This can be statistically inefficient, since parameter sharing can act as a powerful form of regularization, and it is also computationally inefficient, since it means that the number of parameters in shallow embedding methods necessarily grows as $O(|\V|)$.
\item
	Shallow embedding also fails to leverage node attributes during encoding. In many large graphs nodes have attribute information (\textit{e.g.}, user profiles on a social network) that is often highly informative with respect to the node's position and role in the graph.
\item
	Shallow embedding methods are inherently {\em transductive} \cite{hamilton2017inductive}, \textit{i.e.}, they can only generate embeddings for nodes that were present during the training phase, and they cannot generate embeddings for previously unseen nodes unless additional rounds of optimization are performed to optimize the embeddings for these nodes. 
	This is highly problematic for evolving graphs, massive graphs that cannot be fully stored in memory, or domains that require generalizing to new graphs after training. 
\end{enumerate}
Recently, a number of approaches have been proposed to address some, or all, of these issues.
These approaches still fall firmly within the encoder-decoder framework outlined in Section \ref{sec:encdec}, but they differ from the shallow embedding methods of Section \ref{sec:directencoding} in that they use a more complex encoders, often based on deep neural networks and which depend more generally on the structure and attributes of the graph.

\subsubsection{Neighborhood autoencoder methods}\label{sec:neighauto}

\begin{figure}
\centering
\includegraphics[width=0.95\textwidth, bb=0 0 560 210]{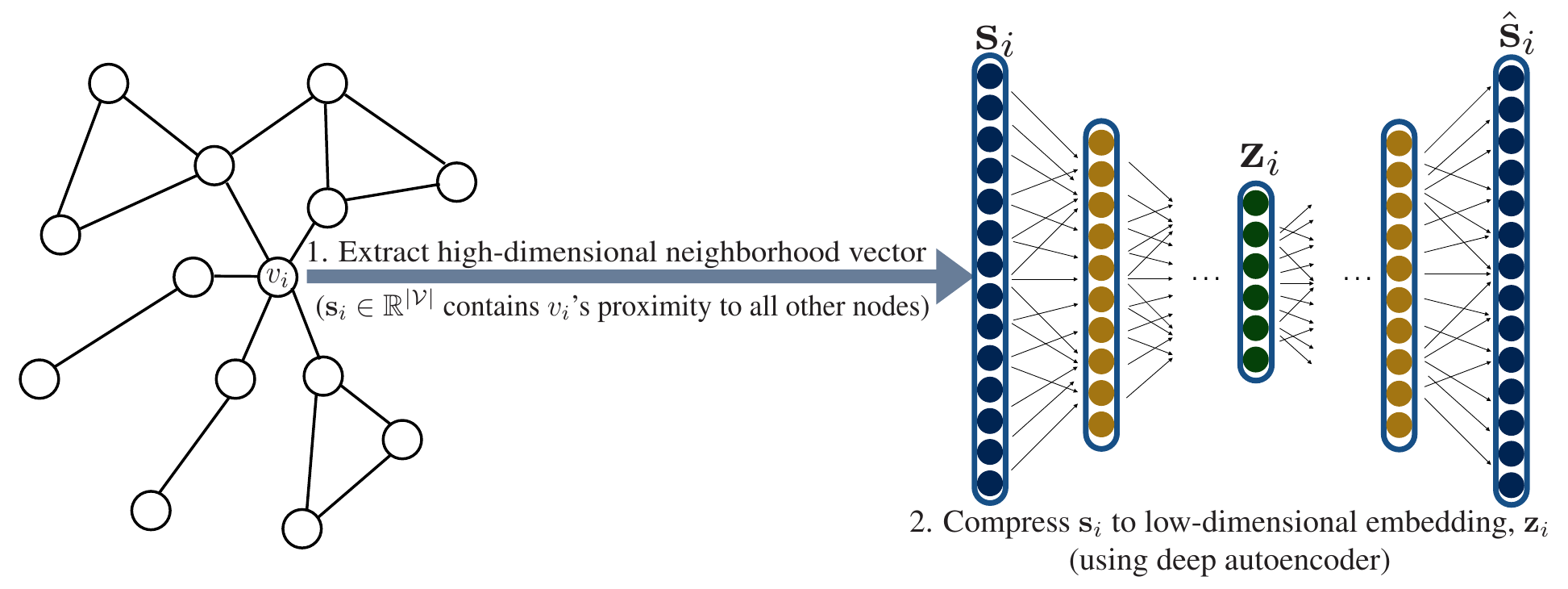}
\caption{
To generate an embedding for a node, $v_i$, the neighborhood autoencoder approaches first extract a high-dimensional neighborhood vector $\mb{s}_i \in \R^{|\V|}$, which summarizes $v_i$'s similarity to all other nodes in the graph.
The $\mb{s}_i$ vector is then fed through a deep autoencoder to reduce its dimensionality, producing the low-dimensional $\mb{z}_i$ embedding. 
}
\label{fig:neighauto}
\end{figure}
Deep Neural Graph Representations (DNGR) \cite{cao2016deep} and Structural Deep Network Embeddings (SDNE) \cite{wang2016structural} address the first problem outlined above: unlike the shallow embedding methods, they directly incorporate graph structure into the encoder algorithm using deep neural networks.
The basic idea behind these approaches is that they use autoencoders---a well known approach for deep learning \cite{hinton2006reducing}---in order to compress information about a node's local neighborhood (Figure \ref{fig:neighauto}). 
DNGR and SDNE also differ from the previously reviewed approaches in that they use a {\em unary decoder} instead of a pairwise one.

In these approaches, each node, $v_i$, is associated with a neighborhood vector, $\mb{s}_i \in \R^{|\V|}$, which corresponds to $v_i$'s row in the matrix $\mb{S}$ (recall that $\mb{S}$ contains pairwise node similarities, \textit{i.e.}, $\mb{S}_{i,j} = s_\G(v_i, v_j)$).
The $\mb{s}_i$ vector contains $v_i$'s similarity with all other nodes in the graph and functions as a high-dimensional vector representation of $v_i$'s neighborhood. 
The autoencoder objective for DNGR and SDNE is to embed nodes using the $\mb{s}_i$ vectors such that the $\mb{s}_i$ vectors can then be reconstructed from these embeddings:
\begin{equation}\label{eq:auto}
\dec(\enc(\mb{s}_i)) = \dec(\mb{z}_i) \approx \mb{s}_i.
\end{equation}
In other words, the loss for these methods takes the following form:
\begin{equation}
\mathcal{L} = \sum_{v_i \in \V}\|\dec(\mb{z}_i) - \mb{s}_i\|_2^2.
\end{equation}
As with the pairwise decoder, we have that the dimension of the $\mb{z}_i$ embeddings is much smaller than $|\V|$ (the dimension of the $\mb{s}_i$ vectors), so the goal is to compress the node's neighborhood information into a low-dimensional vector. 
For both SDNE and DNGR, the encoder and decoder functions consist of multiple stacked neural network layers: each layer of the encoder reduces the dimensionality of its input, and each layer of the decoder increases the dimensionality of its input (Figure \ref{fig:neighauto}; see \cite{hinton2006reducing} for an overview of deep autoencoders). 

SDNE and DNGR differ in the similarity functions they use to construct the neighborhood vectors $\mb{s}_i$ and also in the exact details of how the autoencoder is optimized. 
DNGR defines $\mb{s}_i$ according to the pointwise mutual information of two nodes co-occurring on random walks, similar to DeepWalk and node2vec. SDNE simply sets $\mb{s}_i \triangleq \mb{A}_{i}$, \textit{i.e.}, equal to $v_i$'s adjacency vector. 
SDNE also combines the autoencoder objective (Equation \ref{eq:auto}) with the Laplacian eigenmaps objective (Equation \ref{eq:le}) \cite{wang2016structural}.

Note that the encoder in Equation \eqref{eq:auto} depends on the input $\mb{s}_i$ vector, which contains information about $v_i$'s local graph neighborhood. 
This dependency allows SDNE and DNGR to incorporate structural information about a node's local neighborhood directly into the encoder as a form of regularization, which is not possible for the shallow embedding approaches (since their encoder depends only on the node id).
However, despite this improvement, the autoencoder approaches still suffer from some serious limitations.
Most prominently, the input dimension to the autoencoder is fixed at $|\V|$, which can be extremely costly and even intractable for graphs with millions of nodes. 
In addition, the structure and size of the autoencoder is fixed, so SDNE and DNGR are strictly transductive and cannot cope with evolving graphs, nor can they generalize across graphs. 

\subsubsection{Neighborhood aggregation and convolutional encoders}\label{sec:conv}

\begin{figure}
\centering
\includegraphics[width=0.95\textwidth]{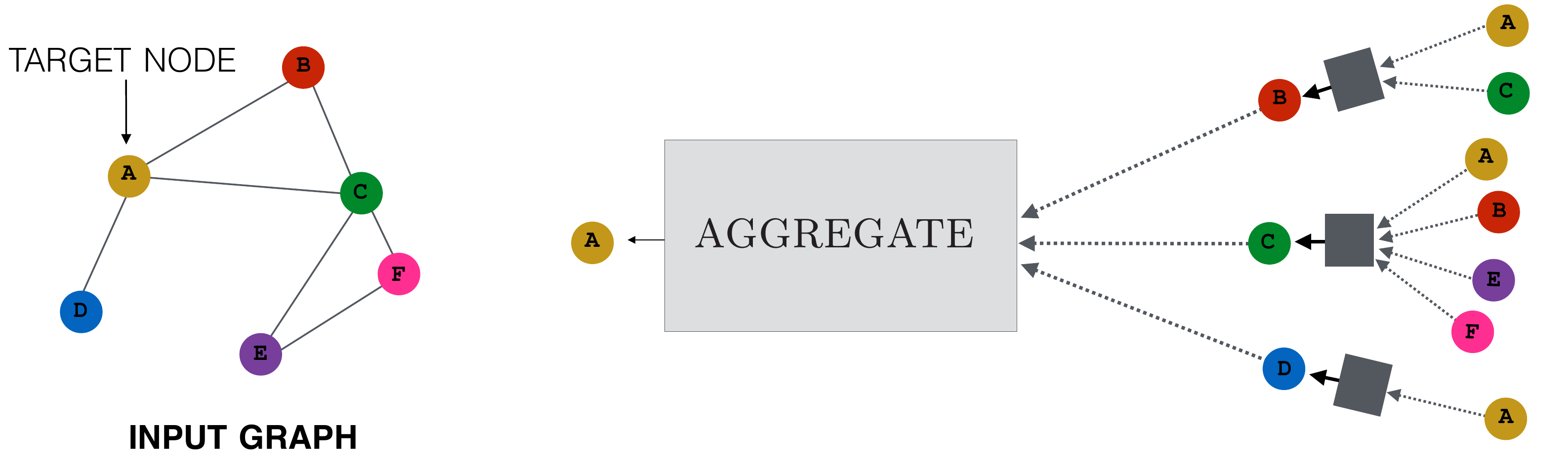}
\vspace{-10pt}
\caption{Overview of encoding in the neighborhood aggregation methods. 
To generate an embedding for node \texttt{A}, the model aggregates messages from \texttt{A}'s local graph neighbors (\textit{i.e.}, \texttt{B}, \texttt{C}, and \texttt{D}), and in turn, the messages coming from these neighbors are based on information aggregated from their respective neighborhoods, and so on.
A ``depth-2'' version of this idea is shown (\textit{i.e.}, information is aggregated from a two-hop neighborhood around node \texttt{A}), but in principle these methods can be of an arbitrary depth.
At the final ``depth'' or ``layer'' the initial messages are based on the input node attributes. 
}
\label{fig:neigh_agg}
\end{figure}

A number of recent node embedding approaches aim to solve the main limitations of the shallow embedding and autoencoder methods by designing encoders that rely on a node's local neighborhood, but not necessarily the entire graph.
The intuition behind these approaches is that they generate embeddings for a node by aggregating information from its local neighborhood (Figure \ref{fig:neigh_agg}). 

Unlike the previously discussed methods, these {\em neighborhood aggregation} algorithms rely on node features or attributes (denoted $\mb{x}_i \in \R^{m}$) to generate embeddings.
For example, a social network might have text data (\textit{e.g.}, profile information), or a protein-protein interaction network might have molecular markers associated with each node.
The neighborhood aggregation methods leverage this attribute information to inform their embeddings. 
In cases where attribute data is not given, these methods can use simple graph statistics as attributes (\textit{e.g.}, node degrees) \cite{hamilton2017inductive}, or assign each node a one-hot indicator vector as an attribute \cite{kipf2016variational,schlichtkrull2017modeling}.
These methods are often called {\em convolutional} because they represent a node as a function of its surrounding neighborhood, in a manner similar to the receptive field of a center-surround convolutional kernel in computer vision \cite{kipf2016semi}.\footnote{These methods also have theoretical connections to approximate spectral kernels on graphs \cite{defferrard2016convolutional}; see \cite{kipf2016semi} for a further discussion.}

\begin{algorithm}[t]
\caption{Neighborhood-aggregation encoder algorithm. Adapted from \cite{hamilton2017inductive}.}
\label{alg:basic}
	\SetKwInOut{Input}{Input}\SetKwInOut{Output}{Output}
    \Input{~Graph $\G(\V,\E)$; input features $\{\mb{x}_v, \forall v\in \V\}$; depth $K$; weight matrices $\{\mb{W}^{k}, \forall k \in [1,K]\}$; non-linearity $\sigma$; differentiable aggregator functions $\{\textsc{aggregate}_k, \forall k \in [1,K]\}$; neighborhood function $\mathcal{N} : v \rightarrow 2^{\V}$}
    \Output{~Vector representations $\mb{z}_v$ for all $v \in \V$}
    \BlankLine
    $\mb{h}^0_v \leftarrow \mb{x}_v, \forall v \in \V$ \;
    \For{$k=1...K$}{
    	  \For{$v \in \V$}{
    	  $\mb{h}^{k}_{\mathcal{N}(v)} \leftarrow \textsc{aggregate}_k(\{\mb{h}_u^{k-1}, \forall u \in \mathcal{N}(v)\})$\;
    	  		$\mb{h}^k_v \leftarrow \sigma\left(\mb{W}^{k}\cdot\textsc{combine}(\mb{h}_v^{k-1}, \mb{h}^{k}_{\mathcal{N}(v)})\right)$
    	  }
    	  $\mb{h}^{k}_v\leftarrow \textsc{normalize}(\mb{h}^{k}_v), \forall v \in \V$
    	}
     $\mb{z}_v\leftarrow \mb{h}^{K}_v, \forall v \in \V$ 
\end{algorithm}

In the encoding phase, the neighborhood aggregation methods build up the representation for a node in an iterative, or recursive, fashion (see Algorithm \ref{alg:basic} for pseudocode).
First, the node embeddings are initialized to be equal to the input node attributes. 
Then at each iteration of the encoder algorithm, nodes aggregate the embeddings of their neighbors, using an aggregation function that operates over sets of vectors.
After this aggregation, every node is assigned a new embedding, equal to its aggregated neighborhood vector combined with its previous embedding from the last iteration.
Finally, this combined embedding is fed through a dense neural network layer and the process repeats. 
As the process iterates, the node embeddings contain information aggregated from further and further reaches of the graph. 
However, the dimensionality of the embeddings remains constrained as the process iterates, so the encoder is forced to compress all the neighborhood information into a low dimensional vector. 
After $K$ iterations the process terminates and the final embedding vectors are output as the node representations.

There are a number of recent approaches that follow the basic procedure outlined in Algorithm \ref{alg:basic}, including graph convolutional networks (GCN)  \cite{kipf2016semi,kipf2016variational,schlichtkrull2017modeling,berg2017graph},
column networks \cite{pham2017column},
and the GraphSAGE algorithm \cite{hamilton2017inductive}. 
The trainable parameters in Algorithm 1---a set of aggregation functions and a set weight matrices $\{\mb{W}^k, \forall k \in [1,K]\}$---specify how to aggregate information from a node's local neighborhood and, unlike the shallow embedding approaches (Section \ref{sec:directencoding}), these parameters are shared across nodes.
The same aggregation function and weight matrices are used to generate embeddings for all nodes, and only the input node attributes and neighborhood structure change depending on which node is being embedded.
This parameter sharing increases efficiency (\textit{i.e.}, the parameter dimensions are independent of the size of the graph), provides regularization, and allows this approach to be used to generate embeddings for nodes that were not observed during training \cite{hamilton2017inductive}.

GraphSAGE, column networks, and the various GCN approaches all follow Algorithm \ref{alg:basic} but differ primarily in how the aggregation (line 4) and vector combination (line 5) are performed.
GraphSAGE uses concatenation in line 5 and permits general aggregation functions; the authors experiment with using the element-wise mean, a max-pooling neural network and LSTMs \cite{hochreiter1997long} as aggregators, and they found the the more complex aggregators, especially the max-pooling neural network, gave significant gains.
GCNs and column networks use a weighted sum in line 5 and a (weighted) element-wise mean in line 4. 

Column networks also add an additional ``interpolation'' term before line 7, setting
\begin{equation}
\mb{h}_v^{k'} = \alpha\mb{h}_v^k + (1-\alpha)\mb{h}_v^{k-1},
\end{equation}
where $\alpha$ is an interpolation weight computed as a non-linear function of $\mb{h}_v^{k-1}$ and $\mb{h}_{\mathcal{N}(v)}^{k-1}$.
This interpolation term allows the model to retain local information as the process iterates (\textit{i.e.}, as $k$ increases and the model integrates information from further reaches of the graph). 

In principle, the GraphSAGE, column network, and GCN encoders can be combined with any of the previously discussed decoders and loss functions, and the entire system can be optimized using SGD.
For example, Hamilton et al.\@ \cite{hamilton2017inductive} use an identical decoder and loss as node2vec, while Kipf et al.\@ \cite{kipf2016variational} use a decoder and loss function similar to the Graph Factorization approach. 

Neighborhood aggregation encoders following Algorithm \ref{alg:basic} have been found to provide consistent gains compared to their shallow embedding counterparts 
on both node classification \cite{hamilton2017inductive,kipf2016semi} and link prediction \cite{berg2017graph,kipf2016variational,schlichtkrull2017modeling} benchmarks.
At a high level, these approaches solve the four main limitations of shallow embeddings, noted at the beginning of Section \ref{sec:genencdec}: they incorporate graph structure into the encoder; they leverage node attributes; their parameter dimension can be made sub-linear in $|\V|$; and they can generate embeddings for nodes that were not present during training. 

\subsection{Incorporating task-specific supervision}\label{sec:supervision}

The basic encoder-decoder framework described thus far is by default unsupervised, \textit{i.e.}, the model is optimized, or trained, over set of node pairs to reconstruct pairwise similarity values $s_\G(v_i, v_j)$, which depend only on the graph $\G$.
However, many node embedding algorithms---especially the neighborhood aggregation approaches presented in Section \ref{sec:conv}---can also incorporate task-specific supervision \cite{hamilton2017inductive,kipf2016semi,schlichtkrull2017modeling,yang2016revisiting}.
In particular, it is common for methods incorporate supervision from node classification tasks in order to learn the embeddings.\footnote{The unsupervised pairwise decoder is already naturally aligned with the link prediction task.}
For simplicity, we discuss the case where nodes have an associated binary classification label, but the approach we describe is easily extended to more complex classification settings. 

Assume that we have a binary classification label, $y_i \in \mathbb{Z}$, associated with each node. 
To learn to map nodes to their labels, we can feed our embedding vectors, $\mb{z}_i$, through a logistic, or sigmoid, function $\hat{y}_i = \sigma(\mb{z}^\top_i\boldsymbol{\theta})$, 
where $\boldsymbol{\theta}$ is a trainable parameter vector.
We can then compute the cross-entropy loss between these predicted class probabilities and the true labels:
\begin{equation}\label{eq:crossent}
\mathcal{L} = \sum_{v_i \in \V}y_i\log(\sigma(\enc(v_i)^\top\boldsymbol{\theta})) + (1-y_i)\log(1-\sigma(\enc(v_i)^\top\boldsymbol{\theta})).
\end{equation}
The gradient computed according to Equation \eqref{eq:crossent} can then be backpropagated through the encoder to optimize its parameters. 
This task-specific supervision can completely replace the reconstruction loss computed using the decoder (\textit{i.e.}, Equation \ref{eq:reconstruction}) \cite{hamilton2017inductive,kipf2016semi}, or it can be included along with the decoder loss \cite{yang2016revisiting}.

\subsection{Extensions to multi-modal graphs}\label{sec:multilayer}

While we have focused on simple, undirected graphs, many real-world graphs have complex multi-modal, or multi-layer, structures (\textit{e.g.}, heterogeneous node and edge types), and a number of works have introduced strategies to cope with this heterogeneity.
\subsubsection{Dealing with different node and edge types}
Many graphs contain different types of nodes and edges.
For example, recommender system graphs consist of two distinct layers---users and content---while many biological networks have a variety of layers, with distinct interactions between them (\textit{e.g.}, diseases, genes, and drugs).

A general strategy for dealing with this issue is to (i) use different encoders for nodes of different types \cite{chang2015heterogeneous} and (ii) extend pairwise decoders with type-specific parameters \cite{nickel2016review,schlichtkrull2017modeling}.
For example, in graphs with varying edge types, the standard inner-product edge decoder (\textit{i.e.}, $\mb{z}_i^\top\mb{z}_j \approx \mb{A}_{i,j}$) can be replaced with a bilinear form \cite{chang2015heterogeneous,nickel2016review,schlichtkrull2017modeling}: 
\begin{equation}\label{eq:edgetypes}
\dec_{\tau}(\mb{z}_i, \mb{z}_j) = \mb{z}^\top\mb{A}_{\tau}\mb{z},
\end{equation}
where $\tau$ indexes a particular edge type and $\mb{A}_\tau$ is a learned parameter specific to edges of type $\tau$.
The matrix, $\mb{A}_\tau$, in Equation \eqref{eq:edgetypes} can be regularized in various ways (\textit{e.g.}, constrained to be diagonal) \cite{schlichtkrull2017modeling}, which can be especially useful when there are a large number of edge types, as in the case for embedding knowledge graphs. 
Indeed, the literature on knowledge-graph completion---where the goal is predict missing relations in knowledge graphs---contains many related techniques for decoding a large number of edge types (\textit{i.e.}, relations)  \cite{nickel2016review}.\footnote{We do not review this literature in detail here, and refer the reader to  Nickel et al.\@ \cite{nickel2016review} for a recent review.}

Recently, Dong et al.\@ \cite{dong2017metapath2vec} also proposed a strategy for sampling random walks from heterogeneous graphs, where the random walks are restricted to only transition between particular types of nodes. 
This approach allows many of the methods in Section \ref{sec:randwalk} to be applied on heterogeneous graphs and is complementary to the idea of including type-specific encoders and decoders.

\subsubsection{Tying node embeddings across layers}\label{sec:ohmnet}

\begin{figure}
\centering
\includegraphics[width=0.85\textwidth, bb=0 0 610 397]{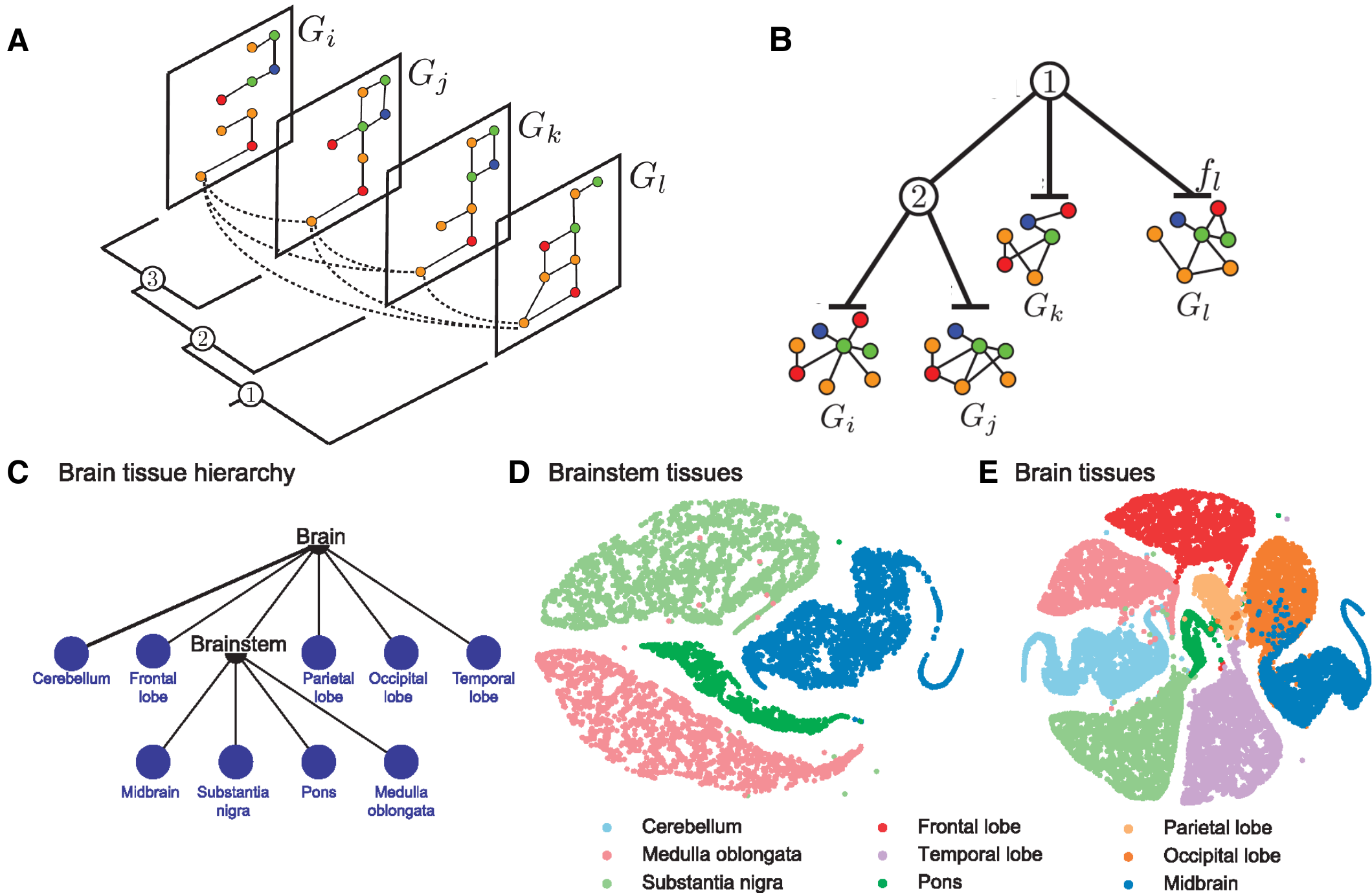}
\caption{\textbf{A}, Example of a 4-layer graph, where the same nodes occur in multiple different layers. This multi-layer structure can be exploited to regularize learning at the different layers by requiring that the embeddings for the same node in different layers are similar to each other. 
\textbf{B}, Multi-layer graphs can exhibit hierarchical structure, where non-root layers in the hierarchy contain the union of the edges present in their child layers---\textit{e.g.}, a biological interaction graph derived from the entire human brain contains the union of the interactions in the frontal and temporal lobes. 
This structure can be exploited by learning embeddings at various levels of the hierarchy, and only applying the regularization between layers that are in a parent-child relationship.
\textbf{C-E}, Example application of multi-layer graph embedding to protein-protein interaction graphs derived from different brain tissues; \textbf{C} shows the hierarchy between the different tissue regions, while \textbf{D} and \textbf{E} visualize the protein embeddings generated at the brainstem and whole-brain layers.
The embeddings were generated using the multi-layer OhmNet method and projected to two dimensions using t-SNE.
Adapted from \cite{zitnik2017tissue}.}
\label{fig:multilayer}
\end{figure}

In some cases graphs have multiple ``layers'' that contain copies of the same nodes (Figure \ref{fig:multilayer}.A). 
For example, in protein-protein interaction networks derived from different tissues (\textit{e.g.}, brain or liver tissue), some proteins occur across multiple tissues.
In these cases it can be beneficial to share information across layers, so that a node's embedding in one layer can be informed by its embedding in other layers.
Zitnik et al.\@ \cite{zitnik2017tissue} offer one solution to this problem, called OhmNet, that combines node2vec with a regularization penalty that ties the embeddings across layers.
In particular, assuming that we have a node $v_i$, which belongs to two distinct layers $\G_1$ and $\G_2$, we can augment the standard embedding loss on this node as follows:
\begin{equation}\label{eq:multireg}
\mathcal{L}(v_i)' = \mathcal{L}(v_i) + \lambda\|\mb{z}_i^{\G_1} - \mb{z}_i^{\G_2}\|
\end{equation}
where $\mathcal{L}$ denotes the usual embedding loss for that node (\textit{e.g.}, from Equation \ref{eq:factloss} or \ref{eq:randloss}), $\lambda$ denotes the regularization strength, and $\mb{z}_i^{\G_1}$ and $\mb{z}_i^{\G_2}$ denote $v_i$'s embeddings in the two different layers, respectively. 

Zitnik et al.\@  further extend this idea by exploiting hierarchies between graph layers (Figure \ref{fig:multilayer}.B). 
For example, in protein-protein interaction graphs derived from various tissues, some layers correspond to interactions throughout large regions (\textit{e.g.}, interactions that occur in any brain tissue) while other interaction graphs are more fine-grained (\textit{e.g.}, only interactions that occur in the frontal lobe). 
To exploit this structure, embeddings can be learned at the various levels of the hierarchy, and the regularization in Equation \eqref{eq:multireg} can recursively applied between layers that have a parent-child relationship in the hierarchy.

\subsection{Embedding structural roles}

So far, all the approaches we have reviewed optimize node embeddings so that nearby nodes in the graph have similar embeddings.
However, in many tasks it is more important to learn representations that correspond to the structural roles of the nodes, independent of their global graph positions (\textit{e.g.}, in communication or transportation networks) \cite{henderson2012rolx}. 
The node2vec approach introduced in Section \ref{sec:randwalk} offers one solution to this problem, as Grover et al.\@ found that biasing the random walks allows their model to better capture structural roles (Figure \ref{fig:n2v_walk}). 
However, more recently, Ribeiro et al.\@ \cite{figueiredo2017struc2vec} and Donnat et al.\@ \cite{donnat2017graphwave} have developed node embedding approaches that are specifically designed to capture structural roles.

Ribeiro et al.\@ propose struc2vec, which involves generating a a series of weighted auxiliary graphs $\G'_k, k=\{1,2,...\}$ from the original graph $\G$, where the auxiliary graph $\G'_k$ captures structural similarities between nodes' $k$-hop neighborhoods.
In particular, letting $R_k(v_i)$ denote the ordered sequence of degrees of the nodes that are exactly $k$-hops away from $v_i$, the edge-weights, $w_k(v_i,v_j)$, in auxiliary graph $G'_k$ are recursively defined as
\begin{equation}
 w_k(v_i, v_j) = w_{k-1}(v_i,v_j) + d(R_{k}(v_i), R_{k}(v_j)),
\end{equation}
where $w_{0}(v_i,v_j) = 0$ and $d(R_{k}(v_i), R_{k}(v_j))$ measures the ``distance'' between the ordered degree sequences $R_{k}(v_i)$ and $R_{k}(v_j)$ (\textit{e.g.}, computed via dynamic time warping \cite{figueiredo2017struc2vec}). 
After computing these weighted auxillary graphs, struc2vec runs biased random walks over them and uses these walks as input to the node2vec optimization algorithm. 

Donnat et al\@. take a very different approach to capturing structural roles, called GraphWave, which relies on spectral graph wavelets and heat kernels \cite{donnat2017graphwave}. 
In brief, we let $\mb{L}$ denote the graph Laplacian---\textit{i.e.}, $\mb{L} = \mb{D} - \mb{A}$ where $\mb{D}$ contains node degrees  on the diagonal and $\mb{A}$ is the adjacency matrix---and we let $\mb{U}$ and $\lambda_i, i=1...|\V|$ denote the eigenvector matrix and eigenvalues of $\mb{L}$, respectively. 
Finally, we assume that we have a heat kernel, $g(\lambda) = e^{-s\lambda}$, with pre-defined scale $s$. 
Using $\mb{U}$ and $g(\lambda)$, GraphWave computes a vector, $\boldsymbol{\psi}_{v_i}$, corresponding to the structural role of node, $v_i \in \V$, as
\begin{equation}\label{eq:wavegraph}
\boldsymbol{\psi}_{v_i} = \mb{U}\mb{G}\mb{U}^\top\mb{v}_i
\end{equation}
where $\mb{G} = \textrm{diag}([g(\lambda_1),...,g(\lambda_{|\V|})])$ and $\mb{v}_i$ is a one-hot indicator vector corresponding to $v_i$'s row/column in the Laplacian.\footnote{Note that Equation \eqref{eq:wavegraph} can be efficiently approximated via Chebyshev polynomials \cite{donnat2017graphwave}.}
Donnat et al\@. show that these $\boldsymbol{\psi}_{v_i}$ vectors implicitly relate to  topological quantities, such as $v_i$'s degree and the number of $k$-cycles $v_i$ is involved in.
They find that---with a proper choice of scale, $s$---WaveGraph is able to effectively capture structural information about a nodes role in a graph (Figure \ref{fig:struct}). 

\begin{figure}[t]
\includegraphics[width=\textwidth, bb=0 0 964 201]{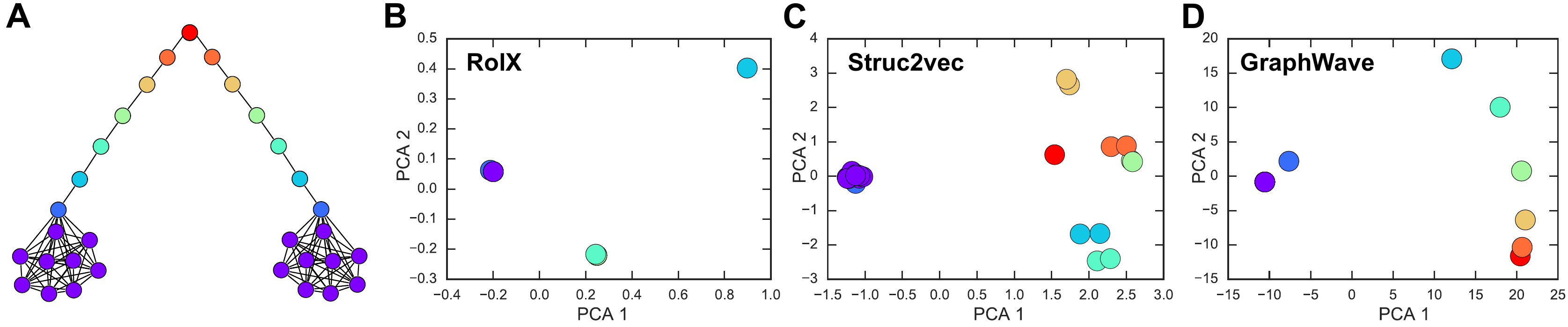}
\caption{
\textbf{A}, Synthetic barbell graph used as a test dataset for detecting structural roles, where nodes are colored according to their structural roles.
In this case, the structural roles (\textit{i.e.}, colors) are computed by examining the degrees of each node's immediate neighbors, and their 2-hop neighbors, and so on (up to $|\V|$-hop neighborhoods). 
 \textbf{B-D}, Visualization of the output of three role-detection algorithms on the barbell graph, where the model outputs are projected using principal components analysis. RolX (\textbf{B}) \cite{henderson2012rolx} is a baseline approach based upon hand-designed features, while struc2vec (\textbf{C}) and GraphWave (\textbf{D}) use different representation learning approaches. 
Note that all methods correctly differentiate the ends of the barbells from the rest of the graph, but only GraphWave is able to correctly differentiate all the various roles. Note also that there are fewer visible nodes in part \textbf{D} compared to \textbf{A} because GraphWave maps identically colored (\textit{i.e.}, structurally equivalent) nodes to the exact same position in the embedding space. 
Reprinted from \cite{donnat2017graphwave}.}
\label{fig:struct}
\end{figure}

\subsection{Applications of node embeddings}\label{sec:nodeapplications}

The most common use cases for node embeddings are for visualization, clustering, node classification, and link prediction, and each of these use cases is relevant to a number of application domains, ranging from computational social science to computational biology. 

\xhdr{Visualization and pattern discovery}
The problem of visualizing graphs in a 2D interface has a long history, with applications throughout data mining, the social sciences, and biology \cite{de2003visual}. 
Node embeddings offer a powerful new paradigm for graph visualization: because nodes are mapped to real-valued vectors, researchers can easily leverage existing, generic techniques for visualization high-dimensional datasets \cite{maaten2008visualizing,tenenbaum2000global}.
For example, node embeddings can be combined with well-known techniques such as t-SNE \cite{maaten2008visualizing} or principal components analysis (PCA) in order to generate 2D visualizations of graphs \cite{perozzi2014deepwalk,tang2015line}, which can be useful for discovering communities and other hidden structures (Figures \ref{fig:karate} and \ref{fig:multilayer}). 

\xhdr{Clustering and community detection}
In a similar vein as visualization, node embeddings are a powerful tool for clustering related nodes, a task that has countless applications from computational biology (\textit{e.g.}, discovering related drugs) to marketing (\textit{e.g.}, discovering related products) \cite{fortunato2010community}. 
Again, because each node is associated with real-valued vector embedding, it is possible to apply any generic clustering algorithm to the set of learned node embeddings (\textit{e.g.}, k-means or DB-scan \cite{ester1996density}).
This offers an open-ended and powerful alternative to traditional community detection techniques, and it also opens up new methodological opportunities, since node embeddings can capture the functional or structural roles played by different nodes, rather than just community structure.

\xhdr{Node classification and semi-supervised learning}
Node classification is perhaps the most common benchmark task used for evaluating node embeddings.
In most cases, the node classification task is a form of semi-supervised learning, where labels are only available for a small proportion of nodes, with the goal being to label the full graph based only on this small initial seed set. 
Common applications of semi-supervised node classification include classifying proteins according to their biological function \cite{grover2016node2vec} and classifying documents, videos, web pages, or individuals into different categories/communities \cite{grover2016node2vec,kipf2016semi,perozzi2014deepwalk,tang2015line}. 
Recently, Hamilton et al.\@ \cite{hamilton2017inductive} introduced the task of inductive node classification, where the goal is to classify nodes that were not seen during training, \textit{e.g.} classifying new documents in evolving information graphs or generalizing to unseen protein-protein interaction networks. 

\xhdr{Link prediction}
Node embeddings are also extremely useful as features for link prediction, where the goal is to predict missing edges, or edges that are likely to form in the future \cite{backstrom2011supervised}. 
Link prediction is at the core of recommender systems and common applications of node embeddings reflect this deep connection, including predicting missing friendship links in social networks \cite{tang2015line} and affinities between users and movies \cite{berg2017graph}. 
Link prediction also has important applications in computational biology. 
Many biological interaction graphs (\textit{e.g.}, between proteins and other proteins, or drugs and diseases) are incomplete, since they rely on data obtained from costly lab experiments. 
Predicting links in these noisy graphs is an important method for automatically expanding biological datasets and for recommending new directions for wet-lab experimentation \cite{lu2003link}. 
More generally, link prediction is closely related to statistical relational learning \cite{getoor2007introduction}, where a common task is to predict missing relations between entities in a knowledge graph \cite{nickel2016review}. 

\section{Embedding subgraphs}\label{sec:subgraph}

\newcommand{\Sub}{\mathcal{S}}

We now turn to the task of representation learning on (sub)graphs, where the goal is to encode a set of nodes and edges into a low-dimensional vector embedding.
More formally, the goal is to learn a continuous vector representation, $\mb{z}_\Sub \in \R^d$, of an induced subgraph $\G[\mathcal{S}]$ of the full graph $\G$, where $\mathcal{S} \subseteq \V$.  Note that these methods can embed both subgraphs ($\mathcal{S} \subset \V$) as well as entire graphs ($\mathcal{S} = \V$).
The embedding, $\mb{z}_\Sub$, can then be used to make predictions about the entire subgraph; for example, one might embed graphs corresponding to different molecules to predict their therapeutic properties \cite{duvenaud2015convolutional}.

Representation learning on subgraphs is closely related to the design of graph kernels, which define a distance measure between subgraphs \cite{vishwanathan2010graph}. 
That said, we omit a detailed discussion of graph kernels, which is a large and rich research area of its own, and refer the reader to \cite{vishwanathan2010graph} for a detailed discussion. 
The methods we review differ from the traditional graph kernel literature primarily in that we seek to learn useful representations from data, rather than pre-specifying feature representations through a kernel function. 

Many of the methods in this section build upon the techniques used to embed individual nodes, introduced in Section \ref{sec:nodes}. 
However, unlike the node embedding setting, most subgraph embedding approaches are fully-supervised, being used for subgraph classification, where the goal is to predict a label associated with a particular subgraph.
Thus, in this section we will focus on the various different approaches for generating the $\mb{z}_\Sub$ embeddings, with the assumption that these embeddings are being fed through a cross-entropy loss function, analogous to Equation \eqref{eq:crossent}.

\subsection{Sets of node embeddings and convolutional approaches}\label{sec:subgraphconv}

There are several subgraph embedding techniques that can be viewed as direct extensions of the convolutional node embedding algorithms (described in Section \ref{sec:conv}).
The basic intuition behind these approaches is that they equate subgraphs with sets of node embeddings.
They use the convolutional neighborhood aggregation idea (\textit{i.e.}, Algorithm 1) to generate embeddings for nodes and then use additional modules to aggregate sets of node embeddings corresponding to subgraphs.
The primary distinction between the different approaches in this section is how they aggregate the set of node embeddings corresponding to a subgraph.

\subsubsection{Sum-based approaches} For example, ``convolutional molecular fingerprints'' introduced by Duvenaud et al.\@ \cite{duvenaud2015convolutional}  represent subgraphs in molecular graph representations by summing all the individual node embeddings in the subgraph:
\begin{equation}\label{eq:subsum}
\mb{z}_{\Sub} = \sum_{v_i \in \Sub}\mb{z}_i,
\end{equation}
where the embeddings, $\{\mb{z}_i, \forall v_i \in \Sub\}$, are generated using a variant of Algorithm \ref{alg:basic}.

Dai et al.\@ \cite{dai2016discriminative} employ an analogous sum-based approach but note that it has conceptual connections to mean-field inference: 
if the nodes in the graph are viewed as latent variables in a graphical model, then Algorithm \ref{alg:basic} can be viewed as a form of mean-field inference where the message-passing operations have been replaced with differentiable neural network alternatives.
Motivated by this connection, Dai et al.\@ \cite{dai2016discriminative} also propose a modified encoder based on Loopy Belief Propagation \cite{murphy1999loopy}.
Using the placeholders and notation from Algorithm 1, the basic idea behind this alternative is to construct intermediate embeddings, $\boldsymbol{\eta}_{i,j}$, corresponding to edges, $(i,j) \in \E$:
\begin{equation}
\boldsymbol{\eta}^{k}_{i,j} = \sigma(\mb{W}^{k}_{\E}\cdot \textsc{combine} (\mb{x}_i, \textsc{aggregate}(\boldsymbol{\eta}_{l,i}^{k-1}, \forall v_l \in \mathcal{N}(v_i) \setminus v_j \})).
\end{equation}
These edge embeddings are then aggregated to form the node embeddings:
\begin{equation}
\mb{z}^{i} = \sigma(\mb{W}^{k}_{\V}\cdot \textsc{combine} (\mb{x}_i, \textsc{aggregate}(\{\boldsymbol{\eta}_{i,l}^{K}, \forall v_l \in \mathcal{N}(v_i)\})).
\end{equation}
Once the embeddings are computed, Dai et al.\@ \cite{dai2016discriminative}, use a simple element-wise sum to combine the node embeddings for a subgraph, as in Equation \eqref{eq:subsum}.

\subsubsection{Graph-coarsening approaches}
Defferrard et al.\@ \cite{defferrard2016convolutional} and Bruna et al.\@ \cite{bruna2013spectral} also employ convolutional approaches, but instead of summing the node embeddings for the whole graph, they stack convolutional and ``graph coarsening'' layers (similar to the HARP approach in Section \ref{sec:randwalk}).
In the graph coarsening layers, nodes are clustered together (using any graph clustering approach), and the clustered node embeddings are combined using element-wise max-pooling.
After clustering, the new coarser graph is again fed through a convolutional encoder and the process repeats. 

Unlike the convolutional approaches discussed in \ref{sec:conv}, Defferrard et al.\@ \cite{defferrard2016convolutional} and Bruna et al.\@ \cite{bruna2013spectral} also place considerable emphasis on designing convolutional encoders based upon the graph Fourier transform \cite{chung1997spectral}.
However, because the graph Fourier transform requires identifying and manipulating the eigenvectors of the graph Laplacian, naive versions of these approaches are necessarily $O(|\V|^3)$. 
State-of-the-art approximations to these spectral approaches (\textit{e.g.} , using Chebyshev polynomials) are conceptually similar to Algorithm \ref{alg:basic}, with some minor variations, and we refer the reader to Bronstein et al. \cite{bronstein2017geometric} for a thorough discussion of these techniques.

\subsubsection{Further variations}
Other variants of the convolutional idea are proposed by Neipert et al.\@ \cite{niepert2016learning} and Kearnes et al.\@ \cite{kearnes2016molecular}. 
Both advocate alternative methods for aggregating sets of node embeddings corresponding to subgraphs:
Kearnes et al.\@ aggregate sets of nodes using ``fuzzy'' histograms instead of a sum, and they also employ edge embedding layers similar to \cite{dai2016discriminative}.
 Neipart et al.\@ define an ordering on the nodes---\textit{e.g.} using a problem specific ordering or by employing an off-the-shelf vertex coloring algorithm---and 
using this ordering, they concatenate the embeddings for all nodes and feed this concatenated vector through a standard convolutional neural network architecture.

\subsection{Graph neural networks}\label{sec:gnns}

In addition to the convolution-inspired embedding approaches discussed above, there is a related---and chronologically prior---line of work on ``graph neural networks'' (GNNs) \cite{scarselli2009graph}.
Conceptually, the GNN idea is closely related to Algorithm \ref{alg:basic}.
However, instead of aggregating information from neighbors, the intuition behind GNNs is that graphs can be viewed as specifying scaffolding for a ``message passing'' algorithm between nodes. 

In the original GNN framework \cite{scarselli2009graph} every node $v_i$ is initialized with a random embedding $\mb{h}_i^{0}$, and at each iteration of the GNN algorithm nodes accumulate inputs from their neighbors according to
\begin{equation}\label{eq:gnn}
\mb{h}_i^k = \sum_{v_j \in \mathcal{N}(v_i)}h(\mb{h}_j, \mb{x}_{i}, \mb{x}_{j}),
\end{equation}
where $h$ is an arbitrary differentiable function of the form $h :  \R^d \times \R^m \times \R^m \rightarrow \R^d$.
Equation \eqref{eq:gnn} is repeatedly applied in a recursive fashion until the embeddings converge, and special care must be taken to ensure that $h$ is a contraction map.
Once the embeddings have converged after $K$ iterations, the final output embeddings are computed as
$\mb{z}_{v_i} = g(\mb{h}^K_i)$, where $g$ is an arbitrary differentiable function of the form
$g : \R^d \rightarrow\R^d$.
Scarselli et al. \cite{scarselli2009graph} discuss various parameterizations of $h$ and $g$ based on multi-layer perceptrons (MLPs), though they are limited by the need to iterate the message passing to convergence and by the restriction that $f$ must be a contraction map.

Li et al.\@ \cite{li2015gated} extend and modify the GNN framework to use Gated Recurrent Units and back propagation through time \cite{cho2014learning}, which removes the need to run Equation \eqref{eq:gnn} to convergence. 
Adapting the GNN framework to use modern recurrent units also allows Li et al.\@ to leverage node attributes for initialization and to use the output of intermediate embeddings of subgraphs. 
In particular, Li et al.'s Gated Graph Neural Networks initialize the $\mb{h}^{0}_i$ vectors using node attributes (\textit{i.e.}, $\mb{h}^0_i = \mb{x}_i)$ and have update equations of the form
\begin{equation}
\mb{h}^k_i = \textrm{GRU}\left(\mb{h}^{k-1}_i, \sum_{v_j \in \mathcal{N}(v_i)}\mb{W}\mb{h}^{k-1}_j\right),
\end{equation}
where $\mb{W} \in \R^{d \times d}$ is a trainable weight matrix and GRU denotes the Gated Recurrent Unit introduced by Cho et al. \cite{cho2014learning}.

Finally, Gilmer et al. \cite{gilmer2017message} discuss another abstraction of GNNs, considering models of the form
\begin{equation}
\mb{h}^k_i = U\left(\mb{h}^{k-1}_i, \sum_{v_j \in \mathcal{N}(v_i)}q(\mb{h}^{k-1}_i, \mb{h}^{k-1}_j)\right),
\end{equation}
where $q : \R^d \times \R^d \rightarrow \R^{d'}$ is a differentiable function that computes the incoming ``messages'' from neighbors and $U : \R^d \times \R^{d'} \rightarrow \R^d$ is a differentiable ``update'' function.
This framework---termed Message Passing Neural Networks (MPNNs)---generalizes Li et al.'s Gated Graph Neural Networks as well as a number of the earlier mentioned convolutional approaches. 
Gilmer et al. discuss a number of variants and extensions of MPNNs (\textit{e.g.}, incorporating edge features) from the perspective of predicting the properties of molecules based on their graph structure. 

All of these graph neural network approaches can in principle be used for node-level embedding tasks, though they are more often used for subgraph-level embeddings. 
To compute subgraph embeddings, any of the aggregation procedures described in Section \ref{sec:subgraphconv} could be employed, but Scarselli et al.\@ \cite{scarselli2009graph} also suggest that the aggregation can be done by introducing a ``dummy'' super-node that is connected to all nodes in the target subgraph.



\subsection{Applications of subgraph embeddings}\label{sec:subapplications}

The primary use case for subgraph embeddings is for subgraph classification, which has important applications in a number of areas.
The most prominent application domain is for classifying the properties of graphs corresponding to different molecules \cite{dai2016discriminative,duvenaud2015convolutional,niepert2016learning,kearnes2016molecular}.
Subgraph embeddings can be used to classify or predict various properties of molecular graphs, including predicting the efficacy of potential solar cell materials \cite{dai2016discriminative}, or predicting the therapeutic effect of candidate drugs \cite{kearnes2016molecular}. 
More generally, subgraph embeddings have been used to classify images (after converting the image to a graph representation) \cite{bruna2013spectral}, to predict whether a computer program satisfies certain formal properties \cite{li2015gated}, and to perform logical reasoning tasks \cite{li2015gated}. 

\section{Conclusion and future directions}

Representation learning approaches for machine learning on graphs offer a power alternative to traditional feature engineering.
In recent years, these approaches have consistently pushed the state of the art on tasks such as node classification and link prediction.
However, much work remains to be done, both in improving the performance of these methods, and---perhaps more importantly---in developing  consistent theoretical frameworks that future innovations can build upon.

\subsection{Challenges to future progress}

In this review, we attempted to unify a number of previous works, but the field as a whole still lacks a consistent theoretical framework---or set of frameworks---that precisely delineate the goals of representation learning on graphs. 
At the moment, the implicit goal of most works is to generate representations that perform well on a particular set of classification or link prediction benchmarks (and perhaps also generate qualitatively pleasing visualizations). 
However, the unchecked proliferation of disparate benchmarks and conceptual models presents a real risk to future progress, and this problem is only exacerbated by the popularity of node and graph embedding techniques across distinct, and somewhat disconnected, subfields within the machine learning and data mining communities. 
Moving forward as a field will require new theoretical work that more precisely describes the kinds of graph structures that we expect the learned representations to encode, how we expect the models to encode this information, and what constraints (if any) should be imposed upon on these learned latent spaces. 

More developed theoretical foundations would not only benefit researchers in the field---\textit{e.g.}, by informing consistent and meaningful benchmark tasks---these foundations would also allow application domain-experts to more effectively choose and differentiate between the various approaches.
Current methods are often evaluated on a variety of distinct benchmarks that emphasize various different graph properties (\textit{e.g.}, community structures, relationship strengths between nodes, or structural roles). 
However, many real-world applications are more focused, and it is not necessary to have representations that are generically useful for a wide variety of tasks. 
As a field, we need to make it clear what method should be used when, and prescribing such use-cases requires a more precise theoretical understanding of what exactly our learned representations are encoding.

\subsection{Important open problems}

In addition to the general challenges outlined above, there are a number of concrete open problems that remain to be addressed within the area of representation learning on graphs.

\xhdr{Scalability}
While most of the works we reviewed are highly scalable in theory (\textit{i.e.}, $O(|\E|)$ training time), there is still significant work to be done in scaling node and graph embedding approaches to truly massive datasets (\textit{e.g.}, billions of nodes and edges). 
For example, most methods rely on training and storing a unique embedding for each individual node.
Moreover, most evaluation setups assume that the attributes, embeddings, and edge lists of all nodes used for both training and testing can fit in main memory---an assumption that is at odds with the reality of most application domains, where graphs are massive, evolving, and often stored in a distributed fashion. 
Developing representation learning frameworks that are truly scalable to realistic production settings is necessary to prevent widening the disconnect between the academic research community and the application consumers of these approaches.  

\xhdr{Decoding higher-order motifs}
While much work in recent years has been dedicated to refining and improving the encoder algorithm used to generate node embeddings, most methods still rely on basic pairwise decoders, which predict pairwise relations between nodes and ignore higher-order graph structures involving more than two nodes.
It is well-known that higher-order structural motifs are essential to the structure and function of complex networks \cite{benson2016higher}, and developing decoding algorithms that are capable of decoding complex motifs is an important direction for future work. 

\xhdr{Modeling dynamic, temporal graphs}
Many application domains involve highly dynamic graphs where timing information is critical---\textit{e.g.}, instant messaging networks or financial transaction graphs. 
However, we lack embedding approaches that can cope with the unique challenges presented by temporal graphs, such as the task of incorporating timing information about edges.
Temporal graphs are becoming an increasingly important object of study \cite{paranjape2017motifs}, and extending graph embedding techniques to operate over them will open up a wide range of exciting application domains. 

\xhdr{Reasoning about large sets of candidate subgraphs}
A major technical limitation of current subgraph embedding approaches is that they require the target subgraphs to be pre-specified before the learning process.
However, many applications seek to {\em discover} subgraphs with certain properties, and these applications require models that can reason over the combinatorially large space of {\em possible} candidate subgraphs.
For example, one might want to discover central subgraphs in a gene regulatory network, or uncover nefarious sub-communities in a social network.  
We need improved subgraph embedding approaches that can efficiently reason over large sets of candidate subgraphs, as such improvements are critical to expand the usefulness of subgraph embeddings beyond the task of basic subgraph classification. 

\xhdr{Improving interpretability}
Representation learning is attractive because it relieves much of the burden of hand designing features, but it also comes at a well-known cost of interpretability. 
We know that embedding-based approaches give state-of-the-art performance, but the fundamental limitations---and possible underlying biases---of these algorithms are relatively unknown. 
In order to move forward, care must be taken to develop new techniques to improve the interpretability of the learned representations, beyond visualization and benchmark evaluation. 
Given the complexities and representational capacities of these approaches, researchers must be ever vigilant to ensure that their methods are truly learning to represent relevant graph information, and not just exploiting statistical tendencies of benchmarks.  

\subsection*{Acknowledgments}
The authors thank Marinka Zitnik, Zoubin Ghahramani, Richard Turner, Stephen Bach, and Manan Ajay Shah for their helpful discussions and comments on early drafts. 
This research has been supported in part by NSF IIS-1149837, DARPA SIMPLEX,
Stanford Data Science Initiative, and Chan Zuckerberg Biohub.
W.L.H. was also supported by the SAP Stanford Graduate Fellowship and an NSERC PGS-D grant. 
The views and conclusions expressed in this material are those of the authors
and should not be interpreted as necessarily representing the official policies or endorsements, either expressed or
implied, of the above funding agencies, corporations, or the U.S. and Canadian governments.

\end{document}